# Surface Plasmon mediated near-field imaging and optical addressing in nanoscience

**A. Drezet, A. Hohenau, J.R. Krenn**
Institute of Physics, Karl-Franzens University Graz, Universitätsplatz 5, A-8010 Graz, Austria

**M. Brun**
Laboratoire d'électronique et de technologie de l'information, CEA-Grenoble
17 rue des Martyrs, 38054, Grenoble cedex 9 France

**S. Huant**
Laboratoire de Spectrométrie Physique, CNRS UMR5588, Université Joseph Fourier Grenoble, BP 87, 38402 Saint Martin d'Hères cedex France

**Abstract**
We present an overview of recent progress in "plasmonics". We focus our study on the observation and excitation of surface plasmon polaritons (SPPs) with optical near-field microscopy. We discuss in particular recent applications of photon scanning tunnelling microscope (PSTM) for imaging of SPP propagating in metal and dielectric wave guides. We show how near-field scanning optical microscopy (NSOM) can be used to optically and actively address remotely nano-objects such as quantum dots. Additionally we compare results obtained with near-field microscopy to those obtained with other optical far-field methods of analysis such as leakage radiation microscopy (LRM).

**1. Introduction**

As is well known conventional optics is diffraction limited to about half of the effective optical wavelength. However the current trend towards miniaturization of optical elements and devices requires methods of observation with high spatial resolutions adapted to the micrometer and submicrometer optical regime. This problem became of particular importance in recent years when intensive investigation of surface plasmon polaritons (SPPs) started. SPPs are electromagnetic waves confined at a dielectric-metal interface. As surface waves, SPPs are exponentially damped in the direction perpendicular to the interface (Raether, 1988). Moreover SPPs allow transfer of optical information in a two dimensional environment and this appealing property can be applied for optical addressing of interacting elements located at a dielectric/metal interface. Therefore SPP optical devices built at such an interface, and including passive nano-structures like mirrors or beam splitter and active elements like molecules or quantum dots are currently under development. Developments such as these raise the prospect of a new branch of photonics using SPPs, for which the term "plasmonics" emerged (Barnes et al., 1997, 1906).

For experimental investigations an important characteristic of these SPP modes is that their spatial extent is governed by the geometry of the nanostructure rather that by the optical wavelength (Krenn et al., 1999). This consequently opens possibilities for breaking the diffraction limit and demands instruments of observation adapted essentially to the subwalength regime and being capable of imaging the propagation of SPPs in their 2D environment. The development of near-field optics over the last 20 years (Courjon, 2003) is thus completely in parallel to the experimental needs of "plasmonics" because near-field microscopy is indeed able to break the diffraction limit imposed by conventional optics. The aim of this article is to present a short overview of recent progress in the field of SPP imaging using essentially optical near-field techniques. Since near-field methods can be used for excitation as well as for collection of SPPs we will present examples of both applications. For this purpose we will here focus principally on results obtained by us or our collaborators in the last few years. In addition, we will discuss other promising techniques of investigations such as leakage radiation microscopy which are far-field methods.

**2. Surface plasmon polariton propagating waves**

Even if experimental indications of the existence of SPPs were already known but not fully understood at the beginning of the 20th century (Wood's anomalies: Wood, 1902; Fano, 1941; Rayleigh, 1907) it was not before the pioneering works of Ritchie (Ritchie, 1957), Otto (Otto, 1968), Kretschmann and Raether (Kretschmann, 1968), that SPPs have received considerable attention experimentally as well as theoretically. It is not however the aim of this article to enter in historical details, neither is it to describe the full range of experiments and applications associated with the discovery of SPPs. For the present purpose we can limit our analysis to the case of a metal film of complex permittivity $\varepsilon_2$ sandwiched between two dielectric media of permittivity $\varepsilon_1$ (substrate) and $\varepsilon_3 < \varepsilon_1$ (superstratum). This system is theoretically simple and to a good extent experimentally accessible. Now since in the broadest sense a SPP is an electron density oscillation bounded at the interface between a metal and an insulator (Raether, 1988; Burke et al., 1986) one can find the dispersion relation of SPP modes propagating in the structure. The analytical implicit SPP dispersion relation is given by:

$$1 + r^p_{1,2} r^p_{2,3} e^{2ik_{z,2}d} = 0 \qquad (1)$$

with the Fresnel coefficients

$$r^p_{i,j} = \frac{k_{z,i}/\varepsilon_i - k_{z,j}/\varepsilon_j}{k_{z,i}/\varepsilon_i + k_{z,j}/\varepsilon_j} \qquad (2)$$

and $k_{z,i} = (\varepsilon_i k_0^2 - k_x^2)^{1/2}$. $k_0$ is the light wavevector in vacuum, $k_x$ (which is in general a complex number $k'_x + ik''_x$) is the SPP wavevector component parallel to the interfaces, and $d$ is the metal film thickness. From this relation one can extract numerically (Hohenau, 2004; Hohenau et al., 2005a,b; Dionne et al., 2005) two relevant SPP modes (both transverse magnetic in character) which are called s mode (symmetric mode) and a mode (antisymmetric mode) respectively (see Fig. 1a). The s mode (a mode) has the maximum of the electromagnetic field on the $\varepsilon_3$-$\varepsilon_2$ ($\varepsilon_1$-$\varepsilon_2$ interface) and the magnetic field on the two interfaces are in phase (counterphase). In the limit of a thick film (i.e. $d$ tending to infinity which in practical means $d \geq 60nm$) one obtains the asymptotic dispersion relation for a semi infinite medium $\varepsilon_2$:

$$k_{x,i} = k_0 \sqrt{\frac{\varepsilon_i \varepsilon_2}{\varepsilon_i + \varepsilon_2}} \qquad (3)$$

with i=1 or 3 corresponding, respectively, to the a and s mode.

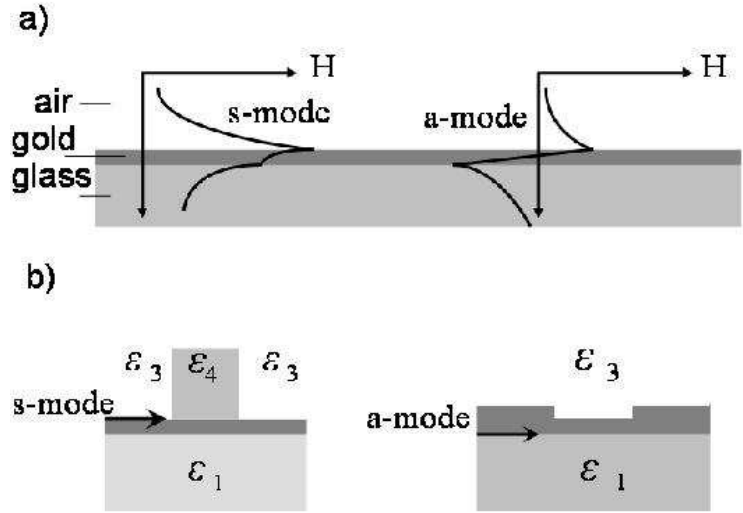

**Fig. 1.** a) The two SPP modes in a layered structure (air/metal/glass) (Raether, 1988). b) The wavevector of the SPP can be affected most efficiently by a super stratum (s-modes) deposited on the metal film, or by a local change of the metal thickness (a-modes).

As shown on Fig.2a an increase of $\varepsilon_3$ leads to a shift to a larger SPP wavevector $k'_x$. This affects the s mode more than the a mode. A similar analysis can be done if we reduce or increase the film thickness (see Fig.2b). In that case the a mode is more affected than the s mode (the wavevector increases when we reduce the thickness). This qualitatively means that local modifications of either $d$ or the permittivity can in complete analogy with conventional 3D optics give us the possibility of modifying the propagation of SPP waves (Fig.1b). In practice this allows the realization of lenses, beam splitters and waveguides for SPPs.

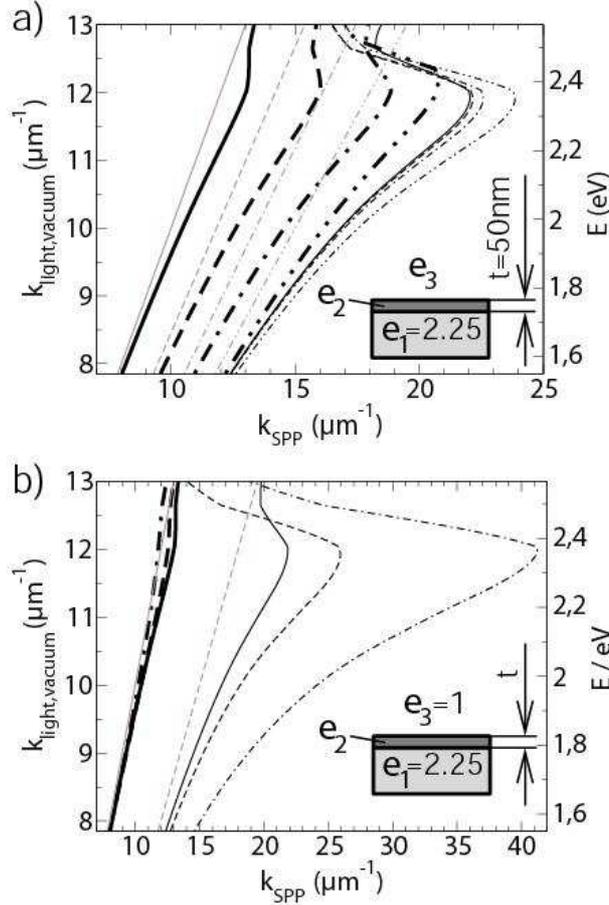

**Fig. 2.** a) Real part of the calculated dispersion relation of the two SPP modes a (thin curves) and s (thick curves) for a t=60 (dashed-dotted curve), 30 (dashed curve) and 15 nm (continuous curve) thin gold film respectively. The curves are calculated using data from (Johnson, 1972). The superstratum is characterized by $\varepsilon_3$= 1 and the substratum by $\varepsilon_1$=2.25. b) The dispersion relation is calculated for various superstratum permittivities $\varepsilon_3$= 1 (continuous curve), 1.4 (dashed curve), 1.8 (dashed-dotted curve), and 2.25 (dashed-double-dotted curve), keeping the gold film thickness constantly equal to t=50 nm. For comparison the light lines for a medium with $\varepsilon$ equal to respectively 1 (continuous gray line), 1.4 (dashed gray line), 1.8 (dashed-dotted gray line), and 2.25 (dashed-double-dotted gray line) are plotted. It should be mentioned that there is also an imaginary part of the wavevector and that the SPP is dominated by an evanescent nature in the region below 3 eV (i.e. the propagation length is below one wavelength) due to the gold interband absorption.

Since any plasmonic devices must be connected to the rest of the optical world for example by optical fibres one must find a way to couple propagating light into SPPs. In order to excite SPPs different methods have been implemented in the past allowing local excitation and thus a study of SPP propagation properties. The Kretschmann method (Kretschmann and Raether, 1968) is based on the use of a glass prism below the sample (air/metal/glass). The interface metal/glass is then illuminated by laser light in total internal configuration (i.e., attenuated total reflection: ATR). Wave-vector matching (Raether, 1988) between light and SPPs at the interface metal/air (s mode) can be achieved if the incidence angle $\theta$ and the glass refractive index $n$ satisfy the relation

$$k_0 n \sin(\theta) = k_0 \, \text{Re}\left\{\sqrt{\frac{\varepsilon_2}{1+\varepsilon_2}}\right\} \qquad (4)$$

Focusing the light on the interface leads to local SPP excitation as sketched on Fig.3a. Another method of SPP excitation consists in optically illuminating a nanostructure located at the top of the air/metal interface (Krenn et al., 2002a). Typically this nanostructure can be a ridge or a particle (Ditlbacher et al., 2002a) lithographically produced at the surface like in Fig.3b but indeed the nanoobject does not need to be in direct contact with the metal film. A single molecule embedded in a PMMA matrix for example can interact with the metal interface during fluorescence emission and the life time is affected by the SPP coupling (Chance et al., 1974, 1975). Another case corresponds to the optical near-field excitation with a near-field scanning optical microscope (NSOM) tip that we will discuss later on. This last method is reminiscent of the original Otto technique (Otto, 1968) using a glass prism with its hypothenuse in quasi contact in front of the metal interface. In this method the subwavelength air gap between the prism and the metal film allows the excitation of SPPs on the top of the film (s mode). Indeed if laser light in total internal reflection impinges on the hypotenuse of the prism the evanescent light couples by the tunnel effect to SPPs and the wave-matching condition is similar to Eq.4 (Raether, 1988). In the case of the nanoobject sketched on Fig.3b (or with a molecule or a NSOM tip (Hecht et al., 1996) the wave-matching is guaranteed by the creation of evanescent modes around the structure considered. Finally it can be observed that the coupling between SPPs propagating at the top of the film and the light travelling through the near-field optical probe is reversible. One can indeed inversely couple the SPP into the near-field tip as exploited in the photon scanning tunnelling microscope (PSTM) techniques discussed in section III. All these possibilities reveal the strong relationship between near-field optics and SPP waves: a relationship due to the the common evanescent character of the electromagnetic fields considered.

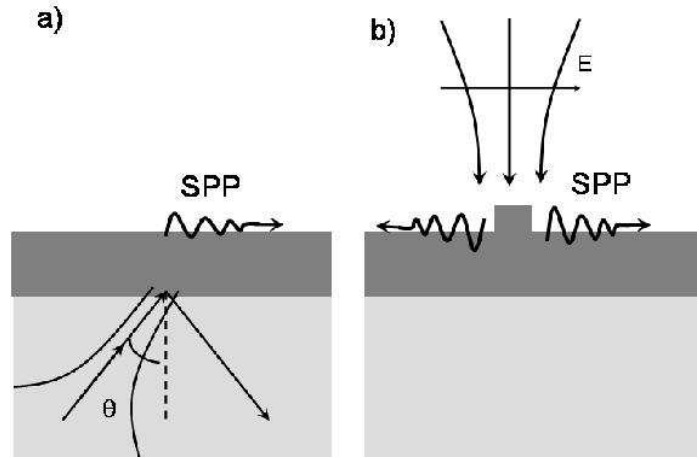

**Fig. 3.** Methods for local excitation of SPPs at the top of a metal film (air/metal/glass). a) Kretschmann configuration (Kretschmann and Raether, 1968) with a focused laser beam in total internal reflection (θ is the incidence angle). b) Local defect method using the coupling between a laser light scattered by a nanostructure (for example ridge, particle) and SPP waves.

## 3. Observation and excitation of SPPs by near-field optical microscopy

### 3.1. Near-field microscopy

The conceptual idea of near-field optics and near-field microscopy came from the seminal proposal of Synge and Einstein (Synge, 1928) to use a subwavelength aperture in a metal film in order to break the diffraction limit imposed by Rayleigh criteria, i.e., the Heisenberg relation $\Delta k_x \Delta x \approx 1$. Indeed classical microscopy is resolution limited due to the fact that diffraction by the objective prohibits the spatial separation of two objects whose distance is typically smaller than about $\lambda/2$ ($\lambda$ being the optical wavelength). Confocal microscopy adapted to the visible can then at best resolve two points separated by 200 nm. However, a nanohole located close to the object can be used to collect or excite evanescent waves (Bethe, 1944; Bouwkamp, 1950; Massey, 1984; Leviathan, 1986; Becker et al., 1981) with wavevectors larger than $2\pi/\lambda$.

This in turn allows a local imaging or excitation in a spatial region size much smaller than $\lambda/2$. This is not in contradiction with Heisenberg relation since the presence of evanescent waves implies typically $\Delta k_x >> 2\pi/\lambda$, which means $\Delta x << \lambda$. Near-field techniques were intensively developed in the mid 1980s with the introduction of local probes inspired by the progress in atomic force microscopy (AFM) and scanning tunnelling microscopy (STM). We refer to (Courjon and Bainier, 1994; Courjon, 2003; Dereux et al., 2000; Girard and Dereux, 1996; Girard, 2005; Greffet and Carminati, 1997) for a complete discussion of the historical developments and for the general theory of near-field microscopy.

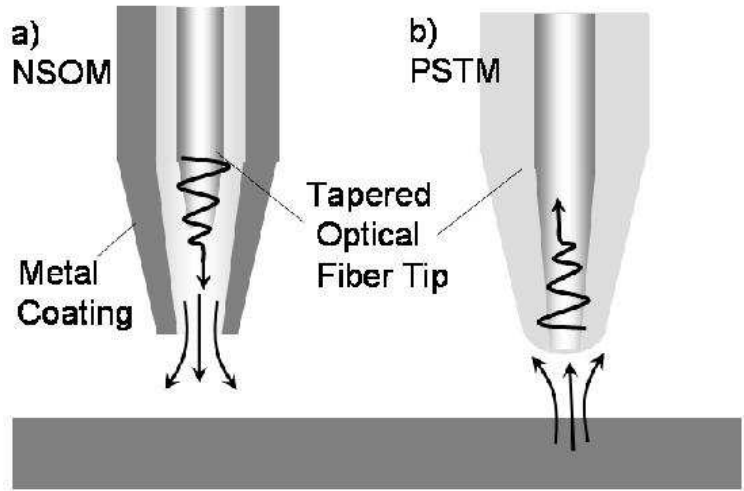

**Fig. 4.** Schematic drawings of a) NSOM and b) PSTM microscopes in respectively excitation or collection modes.

There are typically three kinds of near-field microscopy which are currently used nowadays: the NSOM with an aperture probe, the PSTM and the 'Apertureless' NSOM. The NSOM (also called SNOM, Scanning near-field optical microscope) with an aperture probe (see Fig.4a) is close to the original proposal of Synge (Synge, 1928) and uses as local probe a coated and tapered optical fibre tip with a nano-aperture at the apex (Pohl et al.; 1994; Lewis et al., 1983; Betzig and Trautmann, 1992; Trautman et al., 1994). This configuration can be used either in transmission (Betzig and Trautmann, 1992; Trautman et al., 1994) or in reflection (Fischer et al., 1988). A different approach (see Fig.4b) considers the same tapered fibre tip without metal coating and is used in the so called photon scanning tunnelling microscopy (PSTM) also called scanning tunnelling optical microscope (STOM) (Reddick et al.., 1989; Courjon et al., 1989; de Fornel et al., 1989). The absence of metal coating implies in principle a lower confinement of the electromagnetic fields at the apex resulting in a slightly lower spatial resolution. This can be overcome by considering completely metalized tips (Fischer et al., 1989) in the so called "apertureless" NSOM. In this configuration (Gleyzes et al., 1995; Kawata, 1997; Zenhausern et al., 1994) the sharp tip is used to scatter the light from the near-field to the far-field. This microscope has been accredited of a very high spatial resolution (Zenhausern et al., 1994) and can be applied to create a strongly localized field at the tip apex. This can be useful, e.g., for analyzing surface enhanced Raman scattering (SERS) (Harschuh et al.; Kneip et al., 1997; Nie and Emory, 1997; Sanchez et al., 1999). However the simplification of the system PSTM allows in turn a simplification in the understanding of the images obtained (Dereux et al.; Girard and Dereux, 1996; Girard, 2005). In practices both the NSOM and PSTM methods can be used for collection as well as for excitation in the optical near-field regime (Courjon, 2003). The design of these different tips and probes is still however an open problem and no universal system has been yet proposed (Courjon, 2003; Courjon and Bainier, 1994; Dereux et al., 2000). In the following we will illustrate examples of SPP applications with either the NSOM with aperture (that we will call for simplicity NSOM) or the PSTM methods and we refer to (Gleyzes et al., 1995; Kawata and Inouye, 1995; Zenhausern et al., 1994; Harschuh et al., 2003; Kneip et al., 1997; Nie and Emory, 1997); Sanchez et al., 1999) for works using "apertureless" NSOM.

**3.2. PSTM imaging of SPPs in metal stripes**

One of the most important problem for nanooptics, and thus for plasmonics, concerns the transmission of electromagnetic signals in SPP waveguides. In this context plasmonics overcomes the standard limitation imposed by diffraction in photonic waveguides since SPPs are indeed naturally confined in a 2D environment. Grooves (Seidel, 2004), stripes (Lamprecht et al., 2001; Barnes et al., 2003; Berini, 2000) chains of particles (Quidant et al., 2004; Nomura et al., 2005) and metal photonic-band-gap geometries (Barnes and Sambles, 2006; Bozhevolnyi et al., 2001) have been studied in this context for their ability to guide or block SPP waves. However other problems like the SPP damping during the propagation must be studied precisely before applications to technology might be realized. For this reason the works presented here are limited to gold and silver films due to the rather low value of the imaginary parts of their dielectric function in the visible spectral range.

In a first step we will discuss the case of metallic SPP wave guides (stripes) lithographically produced on a substrate of $SiO_2$. This problem has been extensively studied in the recent years (Weeber et al., 2001, 2003, 2004, 2005; Krenn and Weeber, 2004) based on PSTM analysis in collection mode. Fig.5a shows the principle of the experiments. In a first step a gold structure is built up on a glass substrate using electron beam lithography (EBL) methods (McCord and Rooks, 1997; Hohenau et al., 2006). A layer (typically less than 100 nm thick) of an electron-sensitive resist (as poly(methymethalcrylate), PMMA) is spin coated on a glass substrate which is covered by a thin layer of indium tin oxide (ITO) providing the ohmic conductivity necessary for the EBL process. The sample pattern is written in the resist using the focused beam of a scanning electron microscope (SEM). After chemical development removing the PMMA in the region exposed by the electron beam a gold (or silver) film is evaporated thermally in high vacuum. Finally the rest of the PMMA is removed chemically (lift-off) leaving the metal structure at the top of the glass substrate.

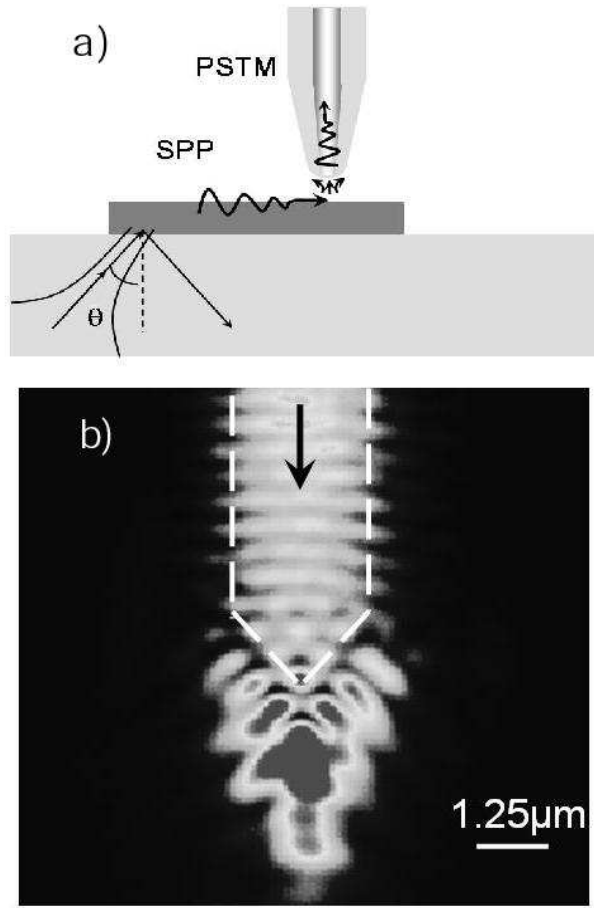

**Fig. 5.** a) Schematic drawing of PSTM in collection mode on a gold structure (stripe) on the top of a glass substrate. θ ≈ 43° is the incidence angle (total internal reflection). b) PSTM image of SPPs launched on the metal/air interface. SPPs propagate in the stripe from top to bottom (3.5 μm width, 60 nm height, $\lambda_0$=633 nm) and are back-reflected at the stripe termination giving rise to the observed interference fringes (Weeber et al., 2001).

During the optical experiments a laser spot is focused on the hypotenuse of the glass prism optically coupled to the sample. As explained previously in the regime of total internal reflection this technique provides a way to excite SPP on top of the metal structure. The typical size of the laser spot at the interface is 10 μm x 10 μm. When this spot is located below the nanostructure SPP waves are launched in the direction of the in-plane component of the wavevector. The optical image is recorded using a purely PSTM probe immersed into the SPP near-field. It has been shown (Weeber et al., 1996; Krenn et al., 1999; (Girard and Dereux, 1996) that the signal collected in such a way by a dielectric tip is directly proportional to the local field intensity at the PSTM tip location. Fig.5b shows an example of such PSTM image. Here SPP waves propagating on a stripe (3.5 μm width, 60 nm height) are imaged using a PSTM at a tip-sample distance below 50 nm. One can clearly see interference fringes caused by the reflection of SPPs at the angular apex of the stripe. In this configuration it is interesting to observe that the fringe visibility $V = (I_{mqx} - I_{min})/(I_{mqx} + I_{min})$ obtained with the PSTM is a direct measure of the real visibility (Bozhevolnyi and Bozhevolnaya, 1999) in spite of the intuitive fact that the PSTM collects at each tip position, information coming from an extended area. This has been experimentally confirmed by recent studies of SPP reflection on Bragg mirrors (Weeber et al., 2004) located on such stripes. In this context further analysis has been made on stripes integrating optical elements such as micro gratings used as Bragg mirrors, or 50-50 beam splitters (Weeber et al., 2004, 2005) (see Sanchez-Gil and Maradudin, 2005) for a recent theoretical analysis of micrograting and Bragg mirrors for SPPs). These elements lithographically produced with EBL (i.e. ridge made gratings) or focused ion beam methods (FIB) (i.e. groove made gratings) could be implemented in a future SPP wave optics based on stripe devices. This is important in particular because SPPs in metal stripe are weakly reflected in the corner of a metal waveguide and continue generally their propagation straight away. If we want to force SPPs to propagate not only in straight line but around corners Bragg mirrors for SPPs (Ditlbacher et al., 2002b; Drezet, 2005b) can be a good solution (in complete analogy with photonic crystals). Other studies (Weeber et al., 2001) show the dependence of the SPP propagation length with the stripe width. One can observe typically a decrease of the propagation length from $L_{SPP}$ = 25 to 4 µm when reducing the stripe width in the interval 54 µm to 70 nm. The upper value corresponds to the limit of the extended gold film. The mechanism justifying this dependence is not clear and could be attributed to the excitation of edge modes inducing more scattering and thus more damping in small width SPP wave guides, or could be connected with the increasing influence of surface roughness when decreasing stripes width. One other possible explanation is that for small stripe width the SPP confinement is not forbidding a direct coupling going from the metal to the glass through the air (see Berini (2000) for a theoretical analysis). Recently a theoretical analysis (Zia et al., 2005) shows however that the damping mechanism could be understood by taking into account the difference between leaky and bound modes of SPP waveguides. In the same way the near-field of SPPs propagating in 50 µm long stripes reveals the modal structure in the transverse direction showing the very strong confinement of SPP in the stripe (Weeber et al., 2001, 2003, 2004). Such studies could be important for exploiting the SPP coupling between stripes, dielectric wave guides or metal nanowires.

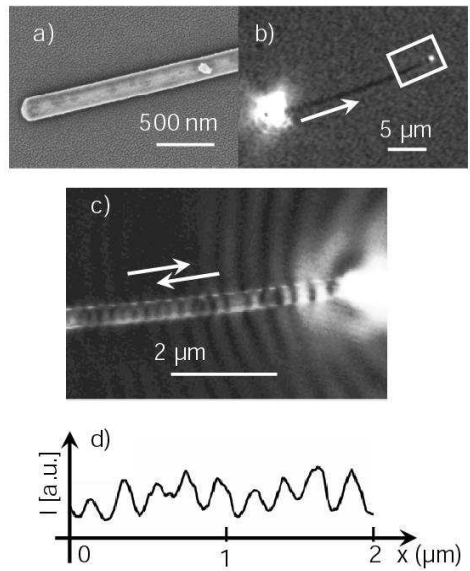

**Fig. 6.** a) SEM image of a silver nanowire (diameter 120 nm) (Ditlbacher et al., 2005). b) Microscopic image showing the illumination spot (Kretschmann configuration) and a less intense spot at the second end of the wire. The arrow shows the direction of SPP propagation. c) The region in the white rectangle is analyzed with a PSTM probe. As indicated by the arrows SPPs are reflected back in the wire when arriving on the exit face (reflectivity ≈ 0.25 (Ditlbacher et al., 2005)). d) Cross-cut of the SPP interference fringes in the wire periodicity=SPP wavelength/2).

Silver nanowires have been similarly extensively studied (Ashley and Emerson, 1974; Dickson and Lyon, 2000; Krenn et al., 2002b; Takahara et al., 1997) in the recent years for applications in waveguiding or optical addressing. Possibilities for passive (Weeber et al., 2004) or active control (Nikolajsen et al., 2004) of SPP propagation in such nanowires or thin stripes have been demonstrated opening the way for integrated

SPP optics. In order to reduce the SPP damping in such structures recent works (Ditlbacher et al., 2005; Graf et al., 2005) analyzed SPP propagation in single-crystalline nanowires. Such structures, having less roughness and thus inducing less scattering and SPP damping, are chemically prepared by a reduction method in aqueous electrolyte solution. The fabrication process leads to cross-section diameters of 13-130 nm and length up to 70 µm. Well separated nanowires can be cast on a glass slide for optical observation. Fig.6a shows the SEM image of a 18.6 µm long nanowire (the picture shows only a zoom) with a diameter of 120 nm. SPPs are excited in the wire using the method of Kretschmann already described. Indeed by moving the laser spot (wavelength

$\lambda_0$ = 785 nm) below one ending of the wire (input) we can excite a propagating SPP along the wire axis. Fig.6b shows a far-field optical image of the sample taken from the top with a microscope objective (60X NA=1.4) and acquired with a charge coupled device (CCD) camera. Light is observed essentially at the input due to scattering. However no light is visible along the nanowire. At the second ending of the wire one can see light emission corresponding to scattering of the outcoming SPP waves. In such a nanowire the SPP wavelength $\lambda_{SPP}$ is much smaller than the laser light and can not couple to far-field light neither in the air nor in the glass side. This shows that one can only analyze SPP propagation in the wire by using a near-field microscope. Fig.6c shows a PSTM image of the wire revealing the presence of a stationary pattern due to SPP reflection at the output face. The periodicity of $\lambda_{SPP}/2$ is found to $L_{SPP}$ = 414 nm which is indeed much smaller than $\lambda_0$ (see Fig.6d). Other far-field analysis made on this kind of structure reveals that the wire behaves like a Fabry-Perot resonator. The single-crystalline nature of the wire is here fundamental for the propagation of SPP waves. It can be indeed observed that electron-beam lithographically produced nanowires are polycrystalline. This apparently induces much more damping which in turn kills the oscillation in the resonator. The value of the SPP propagation length in the single-crystalline $\lambda_{SPP} \approx 10$ µm is well above the values observed for electron-beam lithographically produced wires with similar diameters (Lamprecht et al., 2001; Krenn et al., 2002a,b). For 200nm wide and 50 nm high lithographically produced gold nanowire locally excited at a light wavelength of 800 nm, we have indeed $L_{SPP} \approx 2.5$ nm (Krenn et al., 2002a,b) only.

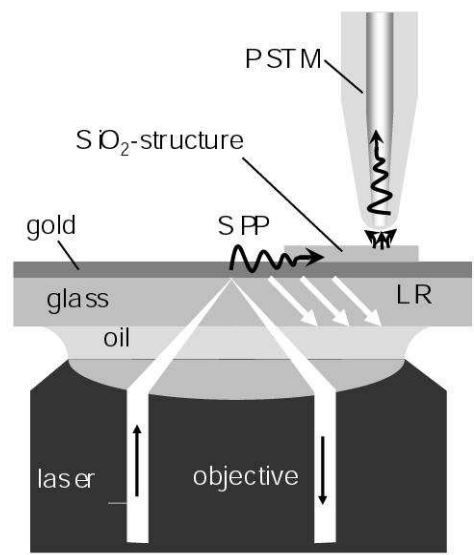

**Fig. 7.** Sketch of a dual LR-PSTM microscope for SPP waves. The immersion objective allows both the excitation of SPPs (in the Kretschman configuration) and the LR far-field recording. The PSTM tip can detect the optical near-field of the propagating SPPs.

**4. PSTM and leakage radiation microscopy**

In addition to near-field microscopy other observation techniques have been developed recently based on far-field optics. Previous works demonstrated (Ditlbacher et al., 2002a; Krenn 2002a) the possibility to record SPP propagation by exploiting fluorescence of dye molecules located on the top of the metal structure. Such methods were in particular applied to study SPP propagation in a 2D Mach-Zehnder interferometer (Ditlbacher et al., 2002b) using Bragg mirror as reflectors. However since molecules have a finite life time

(optical bleaching) these constitute handicaps for quantitative analysis. Additionally in order to prohibit quenching (Moerner and Orrit, 1999) the molecules must be inserted in a matrix film ( such as PMMA at the top of the structure) which modifies strongly the SPP wavevector. For all these reasons investigation are now oriented in the direction of leakage radiation microscopy (Hecht et al., 1996; Bouhelier et al., 2001). This method is based on the fact that in an asymmetric environment (air/metal/glass) SPPs (a mode) can radiate into leaky electromagnetic waves in the glass substrate. In this configuration (which is the counterpart of the Kretschman excitation method) the wavevector matching implies that leakage radiation (LR) can only exit the metal film at a given angle $\theta_{LR}$ such that:

$$k_0 n \sin(\theta_{LR}) = k_0 \, \mathrm{Re}\left\{\sqrt{\frac{\varepsilon_{metal}}{1+\varepsilon_{metal}}}\right\}$$

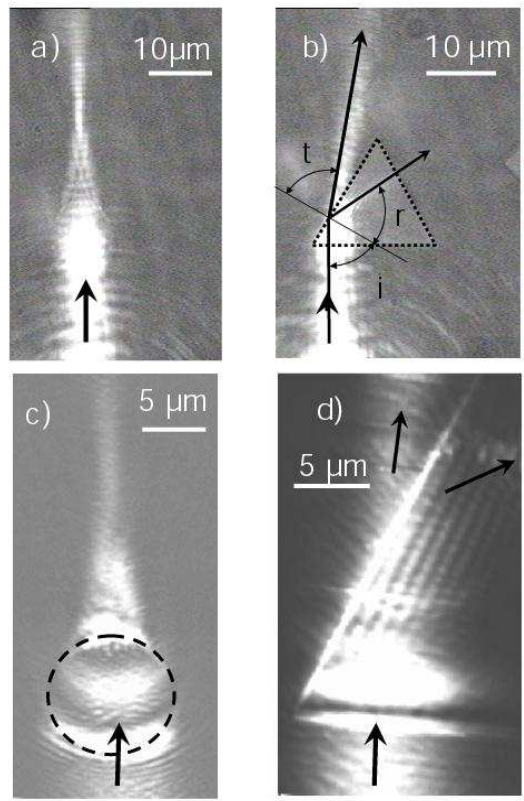

**Fig. 8.** a) LR image of SPPs launched on the top of a gold film (Kretschman configuration) and interacting with a circular $SiO_2$ optical element acting as a lens: SPP are clearly focussed. b) LR image of SPPs launched on the top of a gold film (Kretschman configuration) and interacting with a triangular $SiO_2$ optical element acting as a prism. The lines indicate the incident, reflected and transmitted SPP beams as well as the normal to the interface to which the angles *i* (incidence), *r = i* (reflected), and *t* (transmitted) refer. c) PSTM image of the lens considered in a). d) PSTM image of the prism considered in b). Fringes are visible parallel to the interface (Hohenau et al., 2005a).

In order to observe LR one can use the dual system represented in Fig.7 which allows the simultaneous recording of both LR and near-field image (PSTM) (Hohenau et al., 2005a). With an immersion objective (60X, NA=1.4) one excites SPPs in the Kretschman configuration at the interface metal/air. SPP can be detected as previously by using a PSTM. Simultaneously one can record optical information coming from leaky modes by using a charge coupled device (CCD) camera. This dual method has been recently applied to study SPP propagation in dielectric nano-elements (Hohenau et al., 2005a). It has indeed been remarked previously that only a few structures (i.e. Bragg mirrors) can be used to deflect a SPP wave around the corner of a metal wave guide. Since however 2D optics require optical elements like focuser or lens to be completely consistent we present now an alternative method which is based on the analogy with the 3D case: the use of dielectric structure to refract and reflect SPP waves. Such dielectric elements, located on top of a metal film on which SPPs propagate, can be used to build-up lenses or waveguides. Theoretically this is possible since (see Fig.2) the dielectric medium on the top of the metal film affects strongly the SPP wavevector $k_x$ (Hohenau, 2004). However it should be remarked that the physics is more complicated than in the 3D case because SPP are not only refracted or reflected at interfaces but can be transformed in out-plane scattered light. As an example of application of this dual microscopy we show in Fig.8a comparison between observations done on the same structures with both LR and PSTM techniques. Figs. 8a and 8c compare images of a 40 nm high $SiO_2$ circular structure on top of a 50 nm high gold film. SPPs are excited in the Kretschmann configuration and propagate through the circular element. Refraction induces focussing which is clearly visible on both LR (Fig.8a) and near-field (Fig.8c) images. Similarly Fig.8b and Fig.8d compare LR and near-field images in the case of a 40 nm high $SiO_2$ triangular structure on top of a 50 nm high gold film. SPPs launched on the gold film as previously interact with the triangular structure acting as a prism. One can distinguish the reflected and refracted beam from the incident one on the LR image (Fig.9b). From similar analysis (Hohenau et al., 2005a) of LR one can observe the regime of total internal reflection and deduce the effective medium index $n_{eff} \approx 1.20 \pm 0.05$. The near-field image (Fig.8d) shows more detail inside of the prism and in particular close to the diagonal interface where fringes resulting from the interference between incident and reflected beam are indeed visible. These comparisons prove that both, far-field and near-field imaging, are reliable methods of investigation for SPP propagations. It should be remarked that the LR method has recently been applied in other system such as SPP interferometers using Bragg mirror reflector (Drezet et al., 2005a) and beam splitter (Stepanov et al., 2005). Quantitative analysis and good agreement with theoretical models have been obtained in all cases. This seems to confirm that even if PSTM analysis is in general necessary for a complete understanding of what is happening during the interaction between SPP waves and optical elements LR microscopy constitutes nevertheless a good way to analyze quantitatively some general features of this interaction. For this reason both methods should be used in parallel.

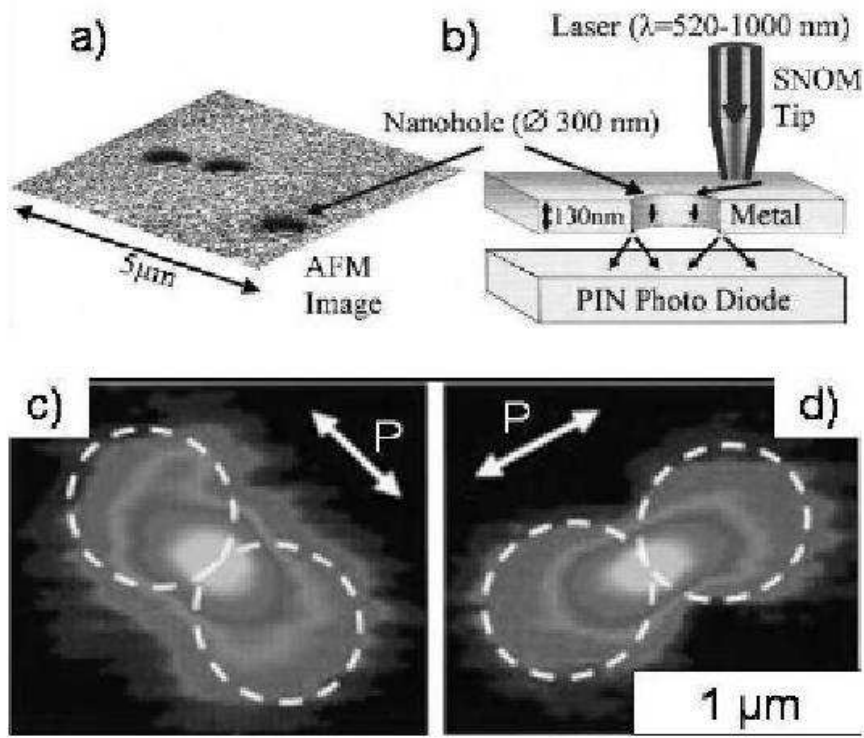

**Fig. 9.** a) Atomic force microscopy (AFM) image of the nano-holes mask discussed in (Sönnichsen et al., 2000). b) Experimental set up for the NSOM excitation of SPP on top of the mask. c)-d) far-field images (recorded with the PIN photodiode sketched in b)) of the SPP waves transmitted by a nanohole: the image is a cartography of the SPP launched by the NSOM tip on the mask. The images are recorded for two orthogonal polarization of the laser.

## 5. SPP addressing of nanoobjects using NSOM

As we explained in section IIIA near-field optical probes can be used to excite and launch SPPs on a metal film because the evanescent field comprises components with finite amplitudes of $k_x$ (see Eqs.1-3). This possibility has been investigated in early works (Hecht et al., 1996) considering the excitation of SPP waves by a NSOM tip (50-100 nm radius aperture) located in front of silver or gold films (60 ± 20 nm thickness). Recent works (Sönnichsen et al., 2003; Brun et al., 2003) showed that a NSOM tip can launch SPPs on gold, silver or aluminum films. Since the laser propagating in the fiber towards the aperture is linearly polarized and since this polarization is a feature conserved at the apex (Betzig and Truatmann, 1992; Drezet, 2001a,b, 2002; Drezet et al., 2004a; Höppener et al., 2002) one can expect indeed a SPP propagation in the direction of the laser polarization. This has been observed in transmission by recording the SPP scattered by a hole located on the metal film (Sönnichsen et al., 2000). The results are represented on Fig.9. SPPs are launched on a 130 nm thick gold film by a NSOM tip (100 nm radius aperture). Nanoholes were in a first step created by spin coating 300 nm polystyrene spheres on a glass substrate before evaporating gold. The spheres together with the metal covering them were then removed by an ultrasonic bath. The nanoholes have a diameter of 300 nm at the end. SPPs scatter into transmitted propagating light after interacting with a nanohole (see Figs.9a,b). This is clearly reminiscent of Ebbesen results obtained with arrays of nanoholes (Ebbesen et al., 1998; Ghaemi et al., 1998) which show an enhanced transmission due to light coupling into SPPs. In the present case (see Figs.9c and d) one can map the SPP intensity on the gold surface by scanning the sample with a NSOM in the close vicinity of a given hole. The hole acts as a test particle cartographying the intensity of the SPP wave launched by the NSOM tip.

This method has been recently applied for addressing selectively nanoobjects (semiconductor quantum dots (QDs)) located below nanoholes on a aluminum film. Such luminescent nanoobjects are self assembled CdTe QDs grown by atomic layer epitaxy (Marsal et al.; Brun et al., 2002). The growth sequence includes a ZnTe buffer layer, a 2.1 nm thick CdTe layer and a protecting 30 nm Znte cap layer. A strong carrier confinement is ensured in ``zero'' dimensional Cd-rich islands (7 nm height, 15 nm diameter). The typical density of such QDs islands is $5 \times 10^{10} cm^{-2}$. The aluminum film is deposited on the top of this semiconductor film containing the QDs. 240 nm diameter nanoholes are created on the top of the metal film by using polystyrene sphere. The QDs can be analyzed spectroscopically by using a NSOM
working at 4.2 K (Brun et al., 2001, 2003, 2004; Matsuda et al., 2003).

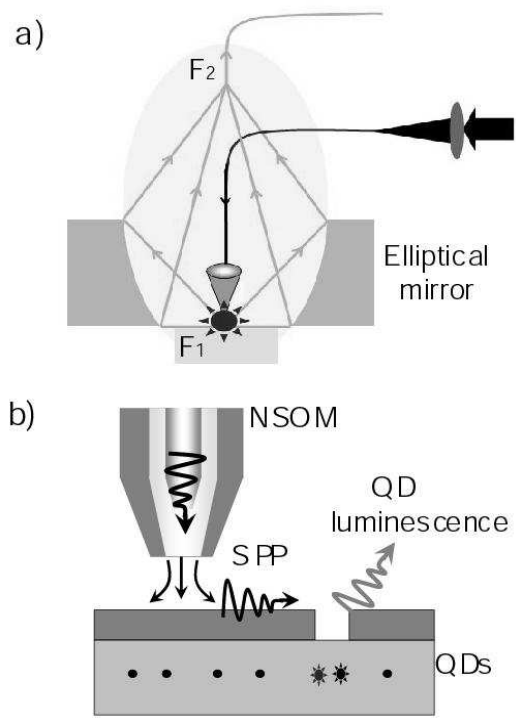

**Fig. 10.** a) Sketch of the cryogenic NSOM microscope used in (Brun et al., 2003). The light emitted in the tip region $F_1$ (corresponding to the first focus of an elliptical mirror) is collected by a multimode fibre located at the second focal point. b) Sketch of the experiment using a NSOM tip to excite SPPs on a aluminum film with nanoholes. An aluminum film containing 240 nm diameter holes covers a CdTe/ZnTe system of QDs whose spectral luminescence induced by SPP excitation is characteristic of a given hole (240 nm). The laser excitation wavelength is 514.5 nm.

Fig.10a shows the principle of a NSOM working in reflection mode at low temperature. The NSOM tip excites locally the sample in the focal region $F_1$ of an elliptical mirror (Brun et al., 2001). The light transmitted by the tip together with the optical information coming from the sample region excited by the NSOM tip are collected by a multimode fibre located at the second focus of the optical mirror. The ensemble is immersed in a cryostat at 4.2 K. The typical experiment is represented in Fig.10b: the NSOM tip launches SPPs on the aluminum film which, after interaction with a hole, excite optically the QDs located below. The collected information shows a specific QDs luminescence spectrum (see Fig.11c). By scanning the sample around the NSOM tip one can realize SPP mapping (compare Fig.9) since the hole act a probe structure for the field emitted by the tip (Hecht et al., 1996; Drezet et al., 2004a,b, 2005a). Quantitative analysis of the total luminescence of the QDs associated with a given hole shows clearly that the QDs excitation is mediated by SPPs propagating on the aluminum film. Fig.11a and Fig.11b show the angular and radial dependence of the intensity. These results agree well with a 2D SPP dipole model supposing an effective dipole located at the tip apex (Hecht et al., 1996; Sönnichsen et al., 2003; Ditlbacher et al., 2002a; Brun et al., 2003; Drezet et al., 2004a,b, 2005a). One finds in particular a SPP propagation length $L_{SPP} \approx 4.2 \mu m$ corresponding to what we know from literature.

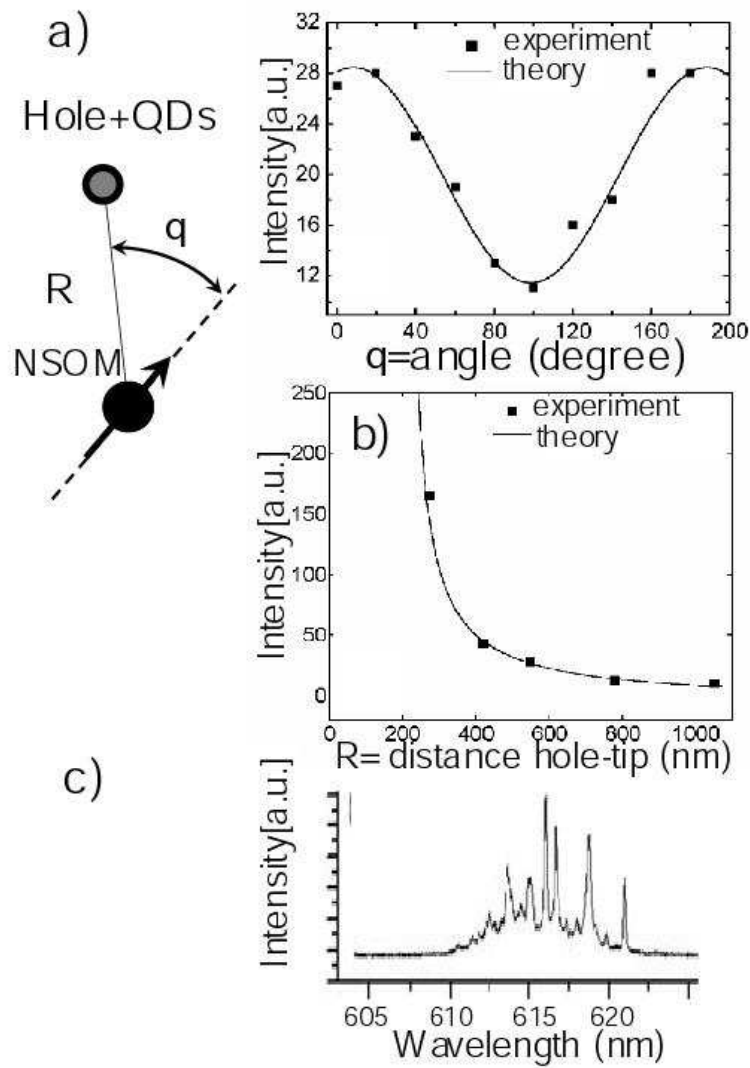

**Fig. 11.** a)-b) Angular and radial mapping of the SPP intensity The SPP intensity is proportional to the QDs luminescence of a given hole. The insert shows the angular and radial parameter $\theta$ and $R$ respectively. c) A typical luminescence spectrum of QDs for a given nano hole. The experiment is made at 4.2 K (Brun et al., 2003, and 2004).

This method of remotely addressing of nano objects is dynamic since by turning the laser polarization one can indeed excite different nano systems. This could have some application in quantum optics where QDs could play the roles of qbits (Salomon et al., 2001; Oliver et al., 2002) and where the generation of quantum entanglement between light and matter is a necessity. Recently some works demonstrated the potentiality of plasmonics to play an important role at the interface between nano and quantum physics (Altewisher et al., 2002; Drezet et al., 2006), however the domain is still for the main part *terra incognita*.

## 6. Conclusion

In this short overview we discussed different optical near-field methods of studying SPPs propagating on a metal/dielectric interface. We showed that for most application the PSTM is an essential tool to understand local properties due to the interaction of SPPs with nano and micro structures. We showed that we can excite SPP with a NSOM probe in order to address dynamically nanoobjects such as QDs which are active elements that promise to play an important role in quantum computation and more generally in quantum physics. Additionally we compared the techniques of investigation based on near-field microscopy with far-field optical methods such as fluorescence and leakage radiation. All these methods are complementary since they enable access to different kind of information, local and global. Finally we conclude by reviewing briefly the topics which were not treated in this article. We didn't study the problem of grating coupling transforming light into SPPs and reciprocally. This topic which is connected with the very beginning of plasmonics (Wood, 1902) has recently been investigated by near-field microscopy. We refer to (Hohenau et al., 2005b; Sanchéz-Gil, 1996) and reference therein for more information on these aspects of plasmonics. These problems are close to those discussed briefly in the article and concerning excitation of SPP modes on periodical arrays of holes (Ebbesen et al., 1998). Such arrays of holes are not only interesting for their enhanced transmission but also for their consequence on wave coherence (Schouten, 2005) and their potential application in quantum optics (Altewisher, 2002). SPP waves have implication in a large area of optics and in particular in imaging where it has been demonstrated that a perfect lens based on materials with a negative index of refraction could recover the optical components associated with subwavelength spatial resolution (Pendry, 2000; Pendry et al. 2004). Applications have been demonstrated in the microwave domain and results in the visible range are expected (Fang et al., 2005; Grigorenko et al., 2005). All these problems concern the near-field regime and certainly require the use of NSOM or PSTM methods for their complete understanding. The same is true for the study of localized surface plasmons in metal nanoparticles and nanoparticle arrays which have been investigated in the last years using optical near-field microscopy (see Krenn et al. 2004a for a review). This reveals once again the intricate relationship existing necessarily between plasmonics and near-field optics and this relation will certainly continue in the next future within the development of nanooptical technology.

## Acknowledgments

For financial support the Austrian Science Foundation and the European Union, under projects FP6 NMP4-CT-2003-505699 and FP6 2002-IST-1-507879 are acknowledged.

## References

Altewisher, E., van Exter, M.P., Voerdman, J.P., 2002. Nature 418, 304.
Ashley, J.C., Emerson, L.C., 1974. Surf. Sci. 41, 6150.
Barnes, W.L., Preist, T.W., Kitson, S.C., Sambles, J.R., 1996. Phys. Rev. B 54, 6227.
Barnes, W.L., Kitson, S.C., Presit, T.W., Sambles, J., 1997. J. Opt. Soc. Am. A 14, 1654.
Barnes, W.L., Dereux, A., Ebbesen, T.W., 2003. Nature 424, 824.
Barnes, W.L., Sambles, J., 2006. Physics World, 17.
Becker, R.S., et al., 1981. Can. J. Phys. 59, 521.
Berini, P., 2000. Phys. Rev. B61, 10484.
Bethe, H.A., 1944, Phys. Rev. 66, 163.
Betzig, E., Trautman, J.K., 1992. Science 257, 189.
Bouhelier, A., et al., 2001. Phys. Rev. B63, 155404.
Bouwkamp, C.J., 1950. Philips Res. Rep. 5, 321.
Bozhevolnyi, S.I., Bozhevolnaya, E.A., 1999. Opt. Lett. 24, 747.
Bozhevolnyi, S.I., Erland, J.E., Leosson, K. Skovgaard, P.M.W., and Hvam, J.M., 2001. Phys. Rev. Lett. 86, 3008.
Brun, M., Huant, S., Woehl, J.C., Motte, J.-F., Marsal, L., Mariette, H., 2001. J. Microsc. 202, 202.
Brun, M., Huant, S., Woehl, J.C., Motte, J.-F., Marsal, L., Mariette, H., 2002. Solid State Commun. 121, 407.

Brun, M., Drezet, A., Mariette, H., chevalier, N., Woehl, J.C., Huant, S., 2003. Europhys. Lett. 64, 634.
Brun, M., et al., 2004. Physica E 21, 219.
Burke, J.J., Stegeman, G.I., Tamir, T., 1986. Phys. Rev. B33, 5186.
Chance, R.R., Prock, A., Silbey, R., 1974. J. Chem. Phys. 60, 2184.
Chance, R.R., Prock, A., Silbey, R., 1975 J. Chem. Phys. 62, 2245.
Courjon, D., Sarayeddine, K., Spajer, M., 1989. Opt. Commun. 71, 23.
Courjon, D., Bainier, C., 1994. Rep. Prog. Phys. 57, 989.
Courjon, D., 2003. Near-field microscopy and near-field optics, Imperial College Press, London.
de Fornel, F., Goudonnet, J.P., Salomon, L., Lesniewska, E., 1989. Proc. SPIE 1139, 77.
Dereux, A., Girard, C., Weeber, J.C., 2000 J. Chem. Phys. 112, 7775.
Dickson, R.M., Lyon, L.A., 2000. J. Phys. Chem. B 104, 6095.
Dionne, J.A., Sweatlock, L.A., Atwater, H.A., Polman, A., 2005. Phys. Rev. B72, 075405.
Ditlbacher, H., Krenn, J.R., Felidj, N., Lamprecht, B., Schider, G., Salemo, M. ,Leitner, A., Aussenegg, F.R., 2002a. Appl. Phys. Lett. 80, 404.
Ditlbacher, H.,Krenn, J.R.,Schider, G.,Leitner, A., Aussenegg, F.R., 2002b. Appl. Phys. Lett. 81, 1762.
Ditlbacher, H., et al., 2005. Phys. Rev. Lett 95, 257403.
Drezet, A., Woehl, J.C., Huant, S., 2001a. J. Microsc. 202, 359.
Drezet, A., Woehl, J.C., Huant, S., 2001b. Europhys. Lett. 54, 736.
Drezet, A., Woehl, J.C., Huant, S., 2002. Phys. Rev. E65, 046611.
Drezet, A., Nasse, M., Huant, S., Woehl, J.C., 2004a. Europhys. Lett. 66, 41.
Drezet, A., Huant, S., Woehl, J.C., 2004b. J. Lumin. 107, 176.
Drezet, A., Brun, M., Woehl, J.C., Huant, S., 2005. J. Korean Phys. Soc. 47, S130.
Drezet, A., Stepanov, A.L., Ditlbacher, H., Hohenau, A., Steinberger, B., Aussenegg, F.R., Leitner, A., Krenn, J.R., 2005. Appl. Phys. Lett. 86, 074104.
Drezet, A., Hohenau, A., Krenn, J.R., 2006. Phys. Rev. A73, 013402.
Ebbesen, T.W., Lezec, H.J., Ghaemi, H.F., Thio, T., Wolff, P.A., 1998. Nature (London) 391, 667.
Fang, N. et al., 2005. Science 308, 534.
Fano, U., 1941, J. Opt. Soc. Am. 31, 213.
Fischer, U.Ch., Dürig, U.T., Pohl, D.W., 1988. Appl. Phys. Lett. 52, 249.
Fischer, U.Ch., Pohl, D.W., 1989. Phys. Rev. Lett. 62, 458.
Ghaemi, H.F., Thio, T., Grupp, D.E., Ebbesen, T.W., Lezec, H.J., 1998. Phys. Rev. B 58, 6779.
Girard, C., Dereux, A., 1996. Rep. Prog. Phys. 59, 657.
Girard, C., 2005 Rep. Prog. Phys. 68, 1883.
Gleyzes, P., Boccara, A.C., Bachelot, R., 1995. Ultramicroscopy 57, 318.
Graf, A., Wagner, D., Ditlbacher, H., Kreibig, U., 2005. Eur. Phys. J. D 34, 263.
Greffet, J.J., Carminati, R., 1997. Prog. Surf. Sci. 56, 133.
Grigorenko, A., et al., 2005. Nature 438, 335.
Hartschuh, A., Sanchez, E.J., Xie, X.S., Novotny, L., 2003. Phys. Rev. Lett. 90, 095503.
Hecht, B., Bielefeldt, H., Novotny, L., Inouye, Y., Pohl, D.W., 1996. Phys. Rev. Lett. 77, 1889.
Hohenau, A., 2004. Phd Thesis: Lichtleitung mit metallischen nanostrukturen, Universität Karl Franzens, Graz Austria.
Hohenau, A. et al., 2005a. Opt. Lett. 30, 893.
Hohenau, A. et al., 2005b Europhys. Lett. 69}, 538.
Hohenau, A., Ditlbacher, H., Lamprecht, B., Krenn, J.R., Leitner, A., Aussenegg, F.R., 2006. to appear in Microelectron. Eng.
Höppener, C., Molenda, D., Fuchs, H., and Naber, A., 2002. Appl. Phys. Lett. 80, 1331.
Johnson, P.B., Christy, R.W., 1972. Phys. Rev. B, 4370.
Kawata, S., Inouye, Y., 1995. Ultramicroscopy 57, 313.
Kneip, K., et al., 1997. Phys. Rev. Lett. 78, 1667.
Kretschmann, E., Raether, H., 1968. Z. Naturforsch. A 23, 2135.
Krenn, J.R., Dereux, A., Weeber, J.C., et al., 1999. Phys. Rev. Lett. 82, 2590.
Krenn, J.R., Ditlbacher, H., Schider, G., Hohenau, A., Leitner, A., Aussenegg, F.R., 2002a. J. Microsc. 209, 167.
Krenn, J.R. et al., 2002b. Europhys. Lett. 60, 663.
Krenn, J.R., Leitner, A., Aussenegg, F.R., 2004. In : Encyclopedia of Nanoscience and Nanotechnology, American Scientific Publishers.
Krenn, J.R., Weeber, J.C., 2004. Philos. Trans. Roy. Soc. Lond. A 326,739.
Lamprecht, B., et al., 2001. Appl. Phys. Lett. 79, 51.
Leviathan, Y., 1986. J. Appl. Phys. 60, 1577.
Lewis, A., Isaacson, M., Muray, A., Harootunian, A., 1983. Biophys. J. 41,405a.

Marsal, L., et al., 2002. J. Appl. Phys. 91, 4936.
Massey, G.A., 1984. Appl. Opt. 23, 658.
Matsuda, K. et al., 2003. Phys. Rev. Lett. 91, 177401.
Moerner, W.E., Orrit, M., 1999. Science 283, 1679.
Nikolajsen, T., Leosson, K., and Bozhevolnyi, S.I., 2004. Appl. Phys. Lett. 85, 5833.
Nie S., Emory, S.R., 1997, Science 275, 1102.
Nomura, W., Ohtsu, M., Yatsui, T., 2005. Appl. Phys. Lett. 86, 181108.
Oliver, W., Yamaguchi, F., Yamamoto, Y., 2002 Phys. Rev. Lett. 88, 037901.
Otto, A., 1968, Z. Phys. 216, 398.
Pohl, D.W., Denk, W., Lanz, M., 1994. Appl. Phys. Lett. 44, 651.
Pendry, J.B., 2000. Phys. Rev. Lett. 85, 3966.
Pendry, J.B., Martin-Moreno, L., Garcia-Vidal, F., 2004. Science 305, 847.
Quidant, R., Girard, C., Weeber, J.C., Dereux, A., 2004. Phys. Rev. B 69, 085407.
Raether, H., 1988. Surface Plasmons, Springer, Berlin.
Rayleigh L., 1907. Proc. Roy. Soc. Lond. Ser. A 79, 399.
Reddick, R.C., Warmack, R.J., Ferrell, T.L., 1989. Phys. Rev. B 39, 767.
Ritchie, R.H., 1957. Phys. Rev. 106, 874.
Salomon, G., Pelton, M., Yamamoto, Y., 2001. Phys. Rev. Lett. 86, 3903.
Sanchez, E.J., Novotny, L., Xie, X.S., 1999. Phys. Rev. Lett. 82, 4014.
Sanchez-Gil, J.A., 1996. Phys. Rev. B 53, 10317.
Sanchez-Gil, J.A., and Maradudin, A.A., 2005. Appl. Phys. Lett. 86, 251106.
Schouten, H.F., 2005. Phys. Rev. Lett. 94, 053901.
Seidel, J., et al., 2004. Phys. Rev. B 69, 121405.
Sönnichsen, C., Duch, A.C. Steiniger, G., Koch, M., vonPlessen, G., Feldmann, J., 2000. Appl. Phys. Lett. 76, 140.
Stepanov, A., et al., 2005. Opt. Lett. 30, 1524.
Synge, E.H., 1928. Phil. Mag. 6, 356.
Takahara, J., Yamagishi, S., Taki, H., Morimoto, A., Kobayashi, T., 1997. Opt. Lett. 22, 475.
Trautman, J.K., Macklin, J.J., Bruns, L.E., Betzig, E., 1994. Nature 369, 40.
Weeber, et al., 1996. Phys. Rev. Lett. 77, 5332.
Weeber, J.C. et al., 2001. Phys. Rev. B 64, 045411.
Weeber, J.C., Lacroute, Y., Dereux, A., 2003. Phys. Rev. B 68, 115401.
Weeber, J.C. et al., 2004. Phys. Rev. B 70, 235406.
Weeber, J.C., Gonzalez, M.U., Baudrion, A.-L., Dereux, A., 2005. Appl. Phys. Lett. 87, 1.
Wood, R.W., 1902. Phil. Mag. 4, 396.
Zenhausern, F., O'Boyle, M.P., Wickramasighe, H.K., 1994. Appl. Phys. Lett. 65, 1623.
Zia, R., Selker, M.D., Brongersma, M.L., 2005. Phys. Rev. B 71, 165431.